\documentstyle[11pt,newpasp,twoside,epsf]{article}

\begin{document}

\title{Nuclear Components in the Bulges of Disk Galaxies}

\author{Marc Balcells, Lilian Dom\'\i nguez-Palmero, Alister Graham}

\affil{Instituto de Astrof\'\i sica de Canarias, 38200 La Laguna, 
Tenerife,
Spain}

\author{ Reynier. F. Peletier}

\affil{School of Physics and Astronomy, University of Nottingham, 
University Park, Nottingham, NG7 2RD, UK}

\begin{abstract}
By combining surface brightness profiles from images taken in the HST/NICMOS F160W and ground-based (GB) $K$ bands, we have obtained NIR profiles
for a well studied sample of inclined disk galaxies, spanning radial ranges
from 20 pc to a few kpc. We fit PSF-convolved Sersic-plus-exponential
laws to the profiles, and compare the results with the fits to the
ground-based data alone.  HST profiles show light excesses over the
best-fit Sersic law in the inner $\sim$1 arcsec.  This is often as a
result of inner power-law cusps similar to the inner profiles
of intermediate-luminosity elliptical galaxies.
\end{abstract}

\section{Introduction}

The nuclear density profile of galaxies is a key tracer of galaxy
formation processes such as dissipation and merging.  Ellipticals may
be divided in two groups according to this profile:  luminous galaxies
with shallow profiles and fainter galaxies with featureless powerlaws
(Faber et al. 1997).  Here we discuss HST profiles of a representative, well-studied sample of galaxy bulges.

\begin{figure}
\plotfiddle{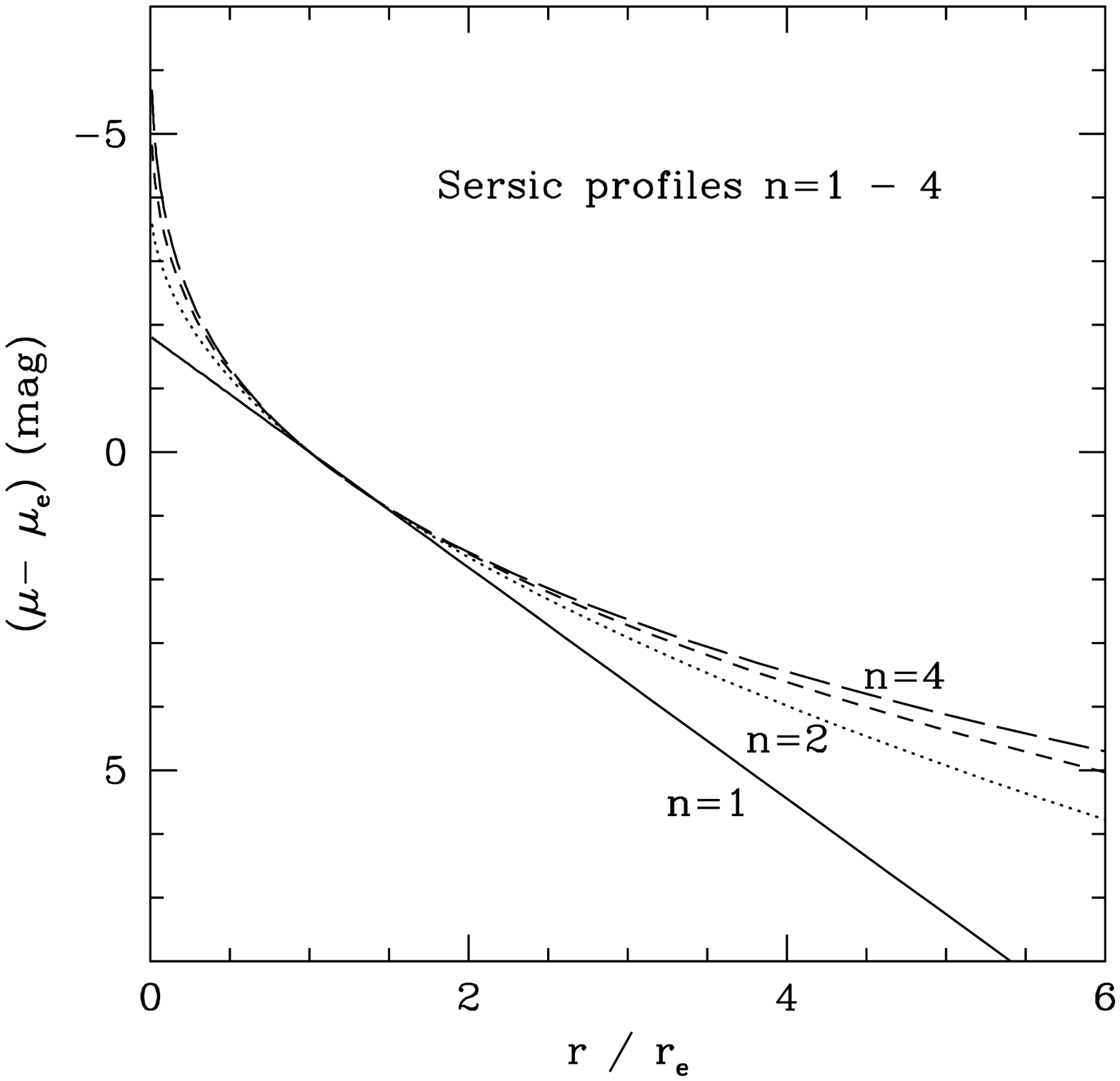}{3cm}{0}{23}{23}{-200}{-90}
\plotfiddle{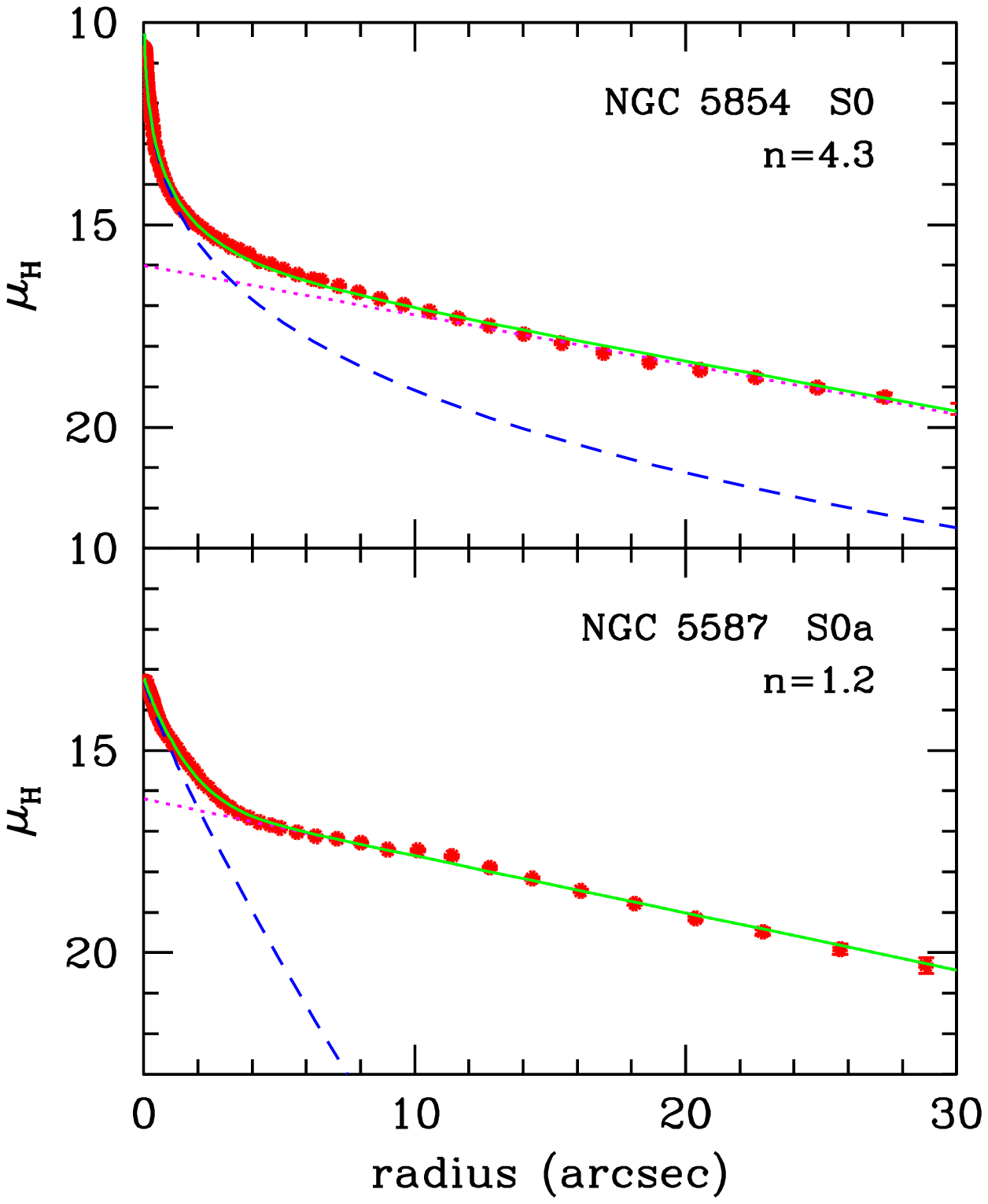}{0cm}{0}{30}{30}{-90}{-100}
\plotfiddle{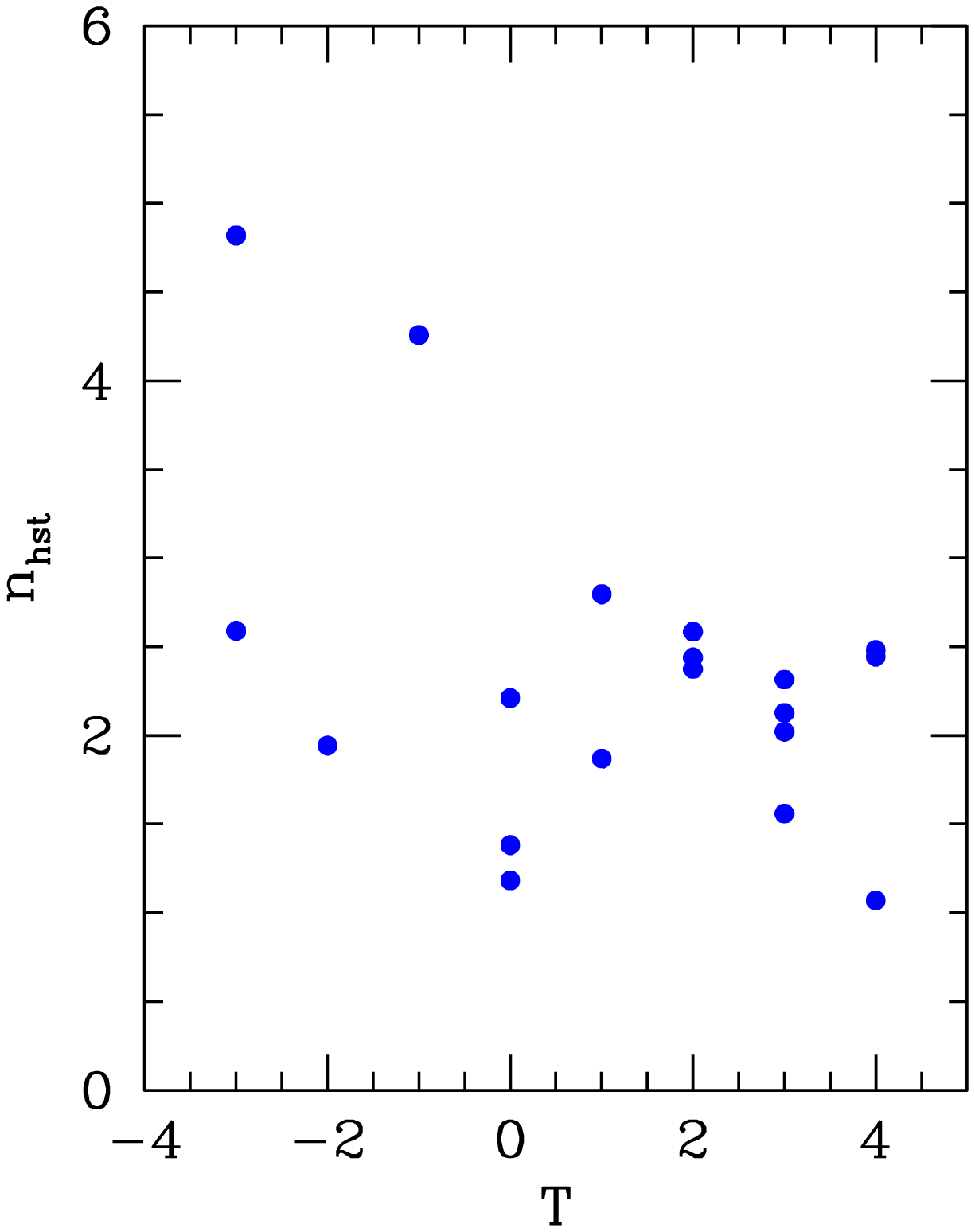}{0cm}{0}{30}{30}{40}{-75}
\caption{
{\it (a) } Four normalized Sersic profiles ($I_{e}=r_{e}=1$)
corresponding to $n=1,2,3,4$.  High-$n$ profiles have both a central
peak and an extended tail.
{\it (b) } Examples of Sersic fits to the bulge $H$-band surface
brightness profiles of two galaxies, from this work. 
{\it (c) } Distribution of shape indices $n_{hst}$ obtained from Sersic law
fits to the HST plus GB $H$-band light profiles.  The abscissa
is T, the morphological type index defined in the RC3 catalog. 
}
\label{MB2:Fig:Sersic}
\end{figure}

Previous work (Andredakis, Peletier, \& Balcells 1995) used ground-based (GB) $K$-band imaging from UKIRT. The surface brightness
profiles of bulges were found to be well described with a family of exponential
power-law profiles given by Sersic's (1968) law:

\[ I(r)=I_{e} \cdot
\exp{\left\{-b_{n}\cdot\left[\left(\frac{r}{r_{e}}\right)^{1/n}-1\right]\right\}},\]

\noindent with $b_{n}\approx 1.9992\cdot n-0.3271$ (Caon, Capaccioli, \& d'Onofrio 
1993).  $R_{e}$ is the effective radius enclosing half the light, 
$I_{e}$ is the intensity at $R_{e}$, and $n$ is the shape index of the profile.  
Figure~\ref{MB2:Fig:Sersic}a shows examples of Sersic profiles,
including the exponential ($n$=1) and de Vaucouleurs ($n$=4) profiles.
Figure~\ref{MB2:Fig:Sersic}b shows examples of Sersic fits to two bulge
profiles.  Andredakis et al.  (1995; see also Graham 2001) found that
the shape index $n$ correlates with the morphological type, or
bulge-to-disk ratio: early-types show de Vaucouleurs ($n\sim 4$)
profiles.  Late-types show ~exponential ($n\sim 1$) profiles.

The goal of this work is to extend the analysis inward toward the
galaxy nucleus, using a combination of HST and ground based imaging to
obtain surface brightness profiles ranging from $\sim$20 pc to a few kpc. 
We perform PSF-convolved Sersic + exponential fits and describe the
behaviour of the profiles in the inner few hundred parsec.

We work with 19 galaxies extracted from the Balcells \& Peletier (1994)
sample of inclined disk galaxies.  For these galaxies, we have obtained
HST/WFPC2 and HST/NICMOS images.  Other analysis of this sample may be
found in Peletier \& Balcells (1996), Peletier et al.\ (1999),
Falc\'on-Barroso, Peletier, \& Balcells (2001).  A downloadable
database of GB images and profiles for the complete Balcells
\& Peletier (1994) sample may be found in Peletier \& Balcells (1997).

\section{Linking HST to Ground-based data}

We analyze the HST/NICMOS F160W NIC2 images (19''$\times$19'', 0.075 
``/pixel) with {\tt galphot} (J{\o}rgensen, Franx, \& 
Kj{\ae}rgaard 1992) and extract elliptically-averaged surface 
brightness profiles.  These are converted to $H$ band magnitudes. We 
then use the elliptically-averaged $K$-band surface brightness 
profiles derived from UKIRT/NICMOS3 images (75''$\times$75'', 
0.291''/pixel) published by Peletier \& Balcells (1997), which we 
transform to $H$ band using the relation $ H-K = 0.111(I-K) - 0.0339 $
(Leitherer et al.\ 1996) and using $I-K$ profiles from Peletier \& 
Balcells (1997).  We then correct the GB profiles from any residual offset using the range $3''\leq r \leq 8''$.  The combined HST plus GB 
profiles are fit with a Sersic bulge plus exponential disk by 
least-squares minimization.  

\section{Inner light profile excesses}

The distribution of shape indices $n$ vs morphological type T is
shown in Figure~\ref{MB2:Fig:Sersic}c.  Albeit with poor statistics, we reproduce the well-known result
that shape indices $n$ reach higher values in early-type galaxies
(Andredakis et al. 1995; Graham 2001).  Intermediate-type galaxies
have $n \sim 1-2$, ie.  the bulges of these galaxies do not follow the
de Vaucouleurs law.  These values of $n$ are systematically higher than those obtained from the fits to the ground-based data alone.  

\begin{figure}
\plotfiddle{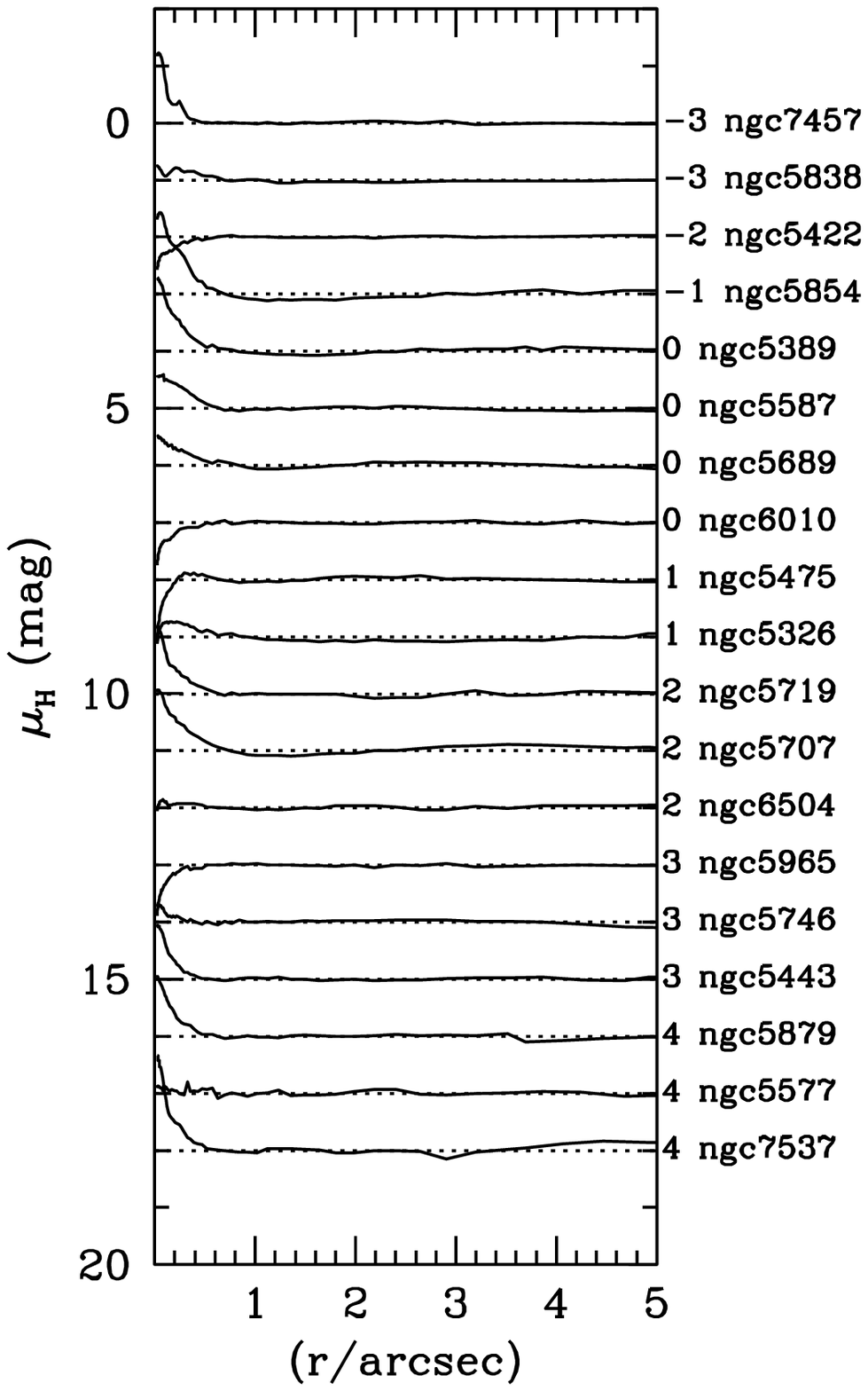}{7cm}{0}{45}{45}{-230}{-110}
\plotfiddle{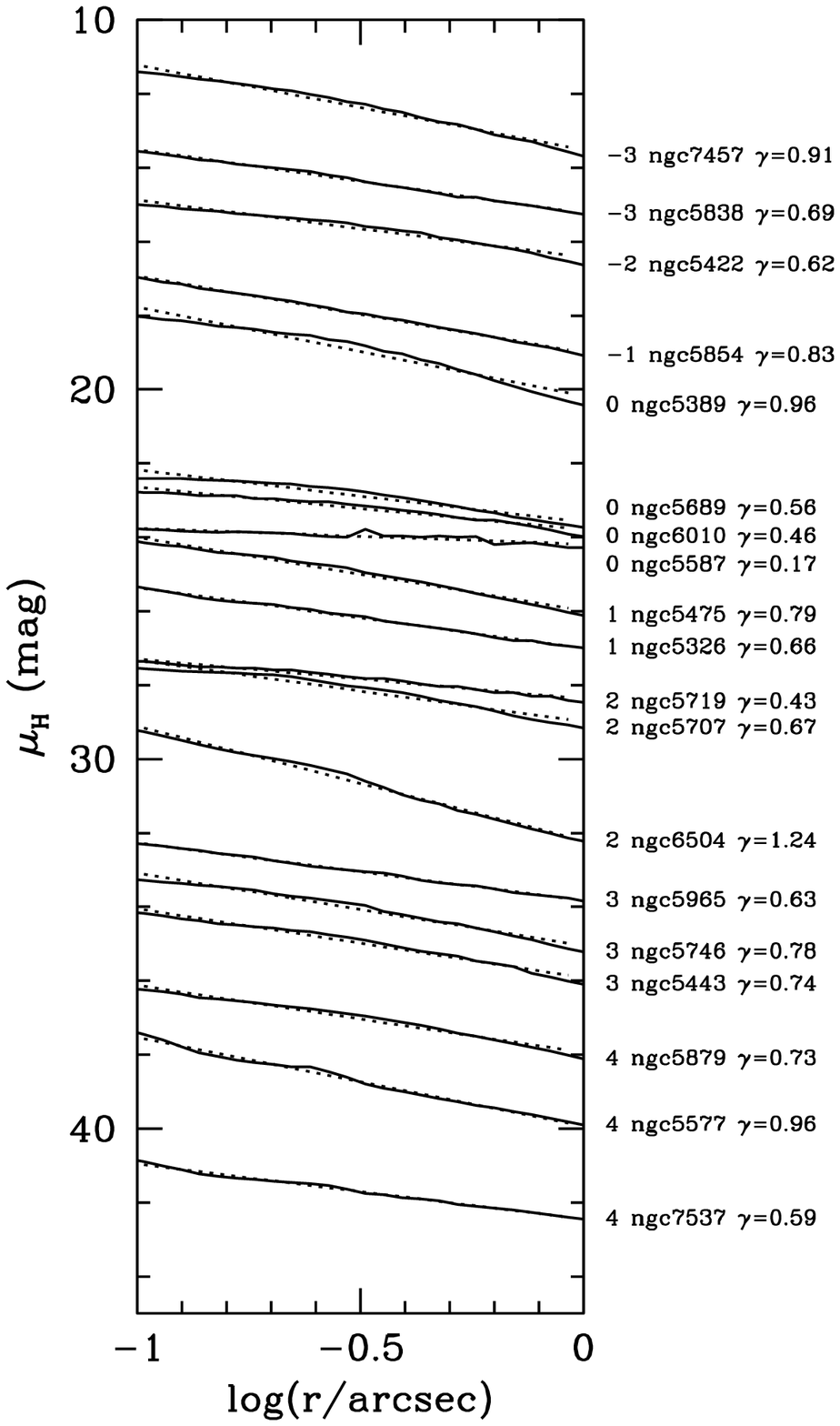}{0cm}{0}{45}{45}{-60}{-70}
\caption{ {\it (a) } Inner residuals of the surface brightness profiles for fits
excluding the central 0.5".  Galaxies are sorted by morphological
type, with earlier types above.  The profiles are stacked with offsets of
1 mag from one to another.  Most galaxies (58\%) show central light
excesses with respect to the Sersic law.
{\it (b) } Bulge surface brightness profiles from 0.1" to 1", against
$\log(r)$.  Profiles are sorted by morphological type, and offset by 1.5
mag from one another for clarity.  The profiles have power-law shapes,
with exponent $\gamma$ given next to each galaxy name.  The dotted lines are
power-law fits to each profile. }
\label{MB2:Fig:Inner}
\end{figure}

Why is that?  Because the bulge light profiles, when seen with HST
spatial resolution, show central light excesses which force the Sersic
fit to a higher $n$.  The fits show deviations in the outer parts.
We show the prevalence of excesses with respect to the Sersic profile
by plotting residuals with respect to the fits performed over a radial range that
excludes the central 0.5" (Fig.~\ref{MB2:Fig:Inner}a). For 58 $\pm$ 11\%  of the sample, we find
central light excesses.  21 $\pm$ 9\% show central depressions.  21\% follow
the Sersic profile over the entire radial range.  The presence of
light excess does not correlate with morphological type in the present data set.

We conclude that the Sersic law cannot provide a full description of
the surface brightness profiles of bulges from $\sim$20pc out to a few
kpc.  Other functional forms are required.  We have investigated Nuker
profiles (Faber et al. 1997).

\section{Inner power-law cusps: Nuker slopes}

Figure~\ref{MB2:Fig:Inner}b shows the bulge surface brightness profiles
for radii $0.1"<r<1"$.  This is the radial range not available to
ground-based observations and revealed by the HST/NICMOS imaging. 
Profiles are well fitted by power laws in this radial range.  We
measure the power law slope index $\gamma$, from the relation
$ I(r) \propto r^{\gamma} $.  Gamma values are given next to each profile.  We obtain values that scatter around $ 0.4 \leq \gamma \leq 1.0 $. 
Only one in 19 profiles has a flat-profile "core", given by a low
$\gamma=0.17$.  The observed range is comparable to that obtained for
intermediate luminosity ellipticals and lenticulars by Nuker fits to
HST data (Faber et al.  1997).  The power-law
profiles continue to rise inward at the limit of HST resolution,
r=0.1".  The value of $\gamma$ does not correlate with morphological type for the present sample. 
The regions inside 1" are typically very reddened, implying
substantial dust obscuration (Peletier et al.\ 1999).  The central stellar
surface brightness may therefore rise even more than represented in
the present profiles.

\section{Conclusions}

(1) When fitting Sersic laws to HST plus ground-based bulge 
    profiles, the shape indices $n$ are higher than those obtained 
    from fits to the GB data alone.  The correlation of $n$ vs 
    T found for GB data remains valid. 
(2) The NIR inner surface brightness profiles of bulges show 
    power-law cusps that continue to rise at the limit of the HST 
    resolution, and have slopes comparable to those of moderate 
    luminosity ellipticals ($0.4\leq\gamma\leq 1.0$).

\end{document}